\documentclass[11pt, a4paper]{article}
\pdfoutput = 1
\usepackage{graphicx}
\usepackage{amsmath}
\begin{document}
 \numberwithin{equation}{section}

\title{String condensation: Nemesis of Black Holes? }

\author{Michael Hewitt \\
      School of Law, Criminal Justice and Computing\\ 
        Canterbury Christ Church University\\
}
      
\date{10 October 2015}
\maketitle

\abstract{This paper puts forward a conjecture that there are no black holes in M theory. We will show that a mechanism to prevent black hole formation is needed in 4 dimensions to make string theory a viable high energy model of quantum gravity. Black hole formation may be averted by a gravity regulation mechanism based on string condensation. In this scenario, black holes are replaced by `hot holograms' that form during gravitational collapse. The geometric conditions based on the properties of free thermalon solutions that are proposed for conversion to a high temperature hologram to occur, however, are local and generic in dimension and could apply throughout M space. This idea can be applied to resolve the problems presented by the process of black hole evaporation, which appears to be inconsistent with quantum information theory.  Whereas, in the conventional view, black holes are real and firewalls are probably a chimera, in the scenario proposed here that situation would be reversed.}

\section{Introduction}

There has been a recent revival of interest in the nature of black holes and debate over the nature of the horizon prompted by the firewall paradox \cite{Mann}.  Four different possibilities for the nature of the collapsed state may be identified with and without horizon, with and without a firewall. The conventional view, which may be extended to the black hole complementarity principle \cite{Susskind93} is that a real horizon forms during collapse, but is not distinguished locally. This is supported by the small magnitude of the Riemannian tidal curvature. However, problems with quantum information during the evaporation process have led to suggestions that the horizon may be surrounded by a locally detectable firewall \cite{Almheiri}\cite{Braunstein}. The fuzzball model \cite{Lunin} avoids the issues associated with horizons and singularities, and these advantages are shared by the proposal presented here that a process of string condensation leads to formation of quasi 2 dimensional hot string regions as end points of gravitational collapse. In this sense, firewalls would be real, but black holes, characterised by a closed horizon, would not. 

\section{Scaling behaviour vs dimension}
The black hole radius $R$ scales as $M^{\frac{1}{d-3}}$, where $d$ is the number of uncompactified spacetime dimensions, so for $d>4$ the excited string states occupy a region larger than a black hole with the same mass, and so can be taken as stable against gravitational collapse. The marginal case is $d=5$, and for $d \leq 4$ a typical excited string state would be hidden by a horizon (see section 4).\\

The gravitational blue shift at a critical acceleration surface is $\propto M^{\frac{1}{d-3}}$ so that the string entropy has the scaling
\begin{displaymath}
S \propto M M^{\frac{1}{d-3}} \propto M^{\frac{d-2}{d-3}} \propto R^{d-2} \propto A
\end{displaymath}
where $A$ measures the critical gravity hypersurface, consistent with a higher dimensional version of the holographic principle \cite{Susskind94}. 

\section{Strings in 4 dimensions}
 An attractive feature of string theory is that it naturally includes gravitons and has finite amplitudes for graviton scattering in the critical dimension. High energy scattering is intermediated by an infinite spectrum of free string states. However, string theory as a perturbative theory of quantum gravity is problematic in 4 dimensions \cite{Hewitt93}, as high energy collisions would form black holes rather than demonstrate string scattering amplitudes or the production of long string states. \\
The generating function \cite{Hewitt93}
\begin{equation}
G(x,y) = \prod_{i=1}^{\infty} \frac{1}{(1-x^i y)(1-x^i y^{-1})} = \prod_{i=1}^{\infty}\frac{1}{1-x^i(y+y^{-1})+x^{2i}} 
= \sum_{n=1,k=1}^{\infty}g_{n,k}x^n y^k
\end{equation}
encodes the distribution of spin against level number. The total squared spin for all states at level $n$ is
\begin{equation}
\sum_{k=1}^{\infty}(g_{n,k} - g_{n, k-1})(2k+1)(k^2 + k)
\end{equation}
or
\begin{equation}
6\sum_{k=1}^{\infty}k^2 g_{n,k} = 3\sum_{k= -\infty}^{\infty}k^2 g_{n,k}
\end{equation}
The operator for squared spin $\hat{J}^2$ on the generating function
\begin{equation}
\hat{J^2} = 3(y \frac{\partial}{\partial y})^2
\end{equation}
and the operator $\hat{n}$ for level number is given by
\begin{equation}
\hat{n} = x \frac{\partial}{\partial x}
\end{equation}
and total values are found by taking the limit $y \rightarrow 1$
On taking this limit, we find
\begin{equation}
\lim_{y \rightarrow 1} \hat{J}^2 G = 3\lim_{y \rightarrow 1} \hat{n} G =6\sum_{n=1}^{\infty}\sigma_{1}x^{n} \lim_{y \rightarrow 1}G
\end{equation}
so that the average squared spin at level $N$ is 3$N$. This implies that the angular momentum of most string states is too small to prevent collapse, and a similar result holds for the electric charge \cite{Hewitt93}. \\
The generating function for eigenstates of $X^2$ is
\begin{equation}
G_R(x,y) = \prod_{n=1}^{\infty}\frac{1}{1-x^n y^{1/n}}p(x)
\end{equation}
where $p(x)$ is the partition function for the other oscillators. Now
\begin{equation}
{\langle R \rangle}^2_N = 3 {\langle X \rangle}^2_N \sim \frac{\pi^2}{\gamma}\sqrt{N}
\end{equation}
where asymptotically
\begin{equation}
\lim_{y \rightarrow 1}G_R \sim \sum_{n=1}^{\infty}q_n x^n
\end{equation}
and
\begin{equation}
q_n \sim \alpha n^{-\beta} \exp (\gamma \sqrt{n})
\end{equation}

It appears that most string states would collapse behind event horizons. However, the fixed nature of the value for the transition temperature calculated from the thermalon winding mode implies that the free spectrum is somehow preserved when gravity is turned on. This supports the possibility of a bag model interpretation for string states in 4 dimensions in analogy to QCD, where quasi-free quarks and gluons are confined to hadrons, which are regions of a different phase to the normal vacuum.\\

In \cite{Hewitt93} it was noted that for heterotic strings, thermal duality implies that there is a maximum of $Z$ against $\beta$ so that the greatest density of strings and information would be around the self-dual temperature, and it was suggested that the heterotic string states in 4 uncompactified dimensions are essentially thin shells. This idea is elaborated below in the context of string condensation, and extended to non-heterotic strings and other dimensions. 

\section{Black hole paradoxes}
While the issue of gravitational collapse for single strings discussed in the previous section applies to 4 dimensions, gravitational collapse presents problems for multi string states in any dimension. To resolve short distances requires concentrating high energy into a small volume - at increasing energy, eventually gravity becomes strong over longer distances, and the resolution deteriorates again. Making gravity consistent with quantum mechanics is problematic for this reason, and macroscopic collapsed objects will also be present for d>4. We therefore expect that a gravitational regulator mechanism should generalise to all regions of M theory and all compactifications, as all will have collapsed states, and so need to avoid paradoxes associated with black holes.
The general problem with black holes in terms of information theory is in getting information out from a photon trap without propagating information faster than light (outside the light cone) and this appears to be inconsistent with the principles of quantum information theory. For example, according to the  Braunstein - Pirandola theorems \cite{Braunstein Pirandola}  the conventional picture of black hole evaporation appears to be inconsistent with unitarity. \\

Theorem 1: A contradiction exists between:
1.a) completely unitarily evaporating black holes,
1.b) a freely falling observer notices nothing special until they pass well within a large black hole's horizon, and 
1.c) the black hole interior Hilbert space dimensionality may be well approximated as the exponential of the Bekenstein Hawking entropy.\\

Theorem 2: A contradiction exists between: 
2.a) completely unitarily evaporating black holes,
2.b) large black holes are described by local physics  (no signals faster than light)    and 
2.c) externally, a large black hole should resemble its classical theoretical counterpart (aside from its slow evaporation).\\

There is also an issue with holographic principle for stellar collapse \cite{Hewitt2015}.The information content of star $\sim 10^{60}$  bits exceeds the holographic limit during collapse inside the horizon before reaching the singularity. String physics is apparently not relevant at this point, in the usual interpretation, so that the conventional collapse process becomes inconsistent with the holographic principle unless information begins to be destroyed before the singularity is reached.\\

\section{The thermalon and hot strings}
The thermalon or thermal scalar \cite{Hewitt2014} \cite{Mertens et al 1} describes a string condensate that forms at the Hagedorn transition \cite{Hagedorn}\cite{AtickWitten}\cite{Sathiapalan}\cite{ObrienTan}.  It is the Euclidean single wrapping mode related to the thermal path integral for a single string. The thermalon is not a normal propagating state but rather a macroscopic order parameter, and provides an effective theory of the thermal string condensate.
The thermalon mode can be interpreted as a deformation of the vacuum \cite{Hewitt2015}, in thermal equilibrium with the normal vacuum. There is a continuous family of solutions, which can be parametrized by warp factor or energy \cite{Hewitt2014}. The thermalon or thermal scalar has an interpretation in terms of a micro-canonical ensemble of strings, as it can be shown to describe their time averaged properties \cite{Mertens et al 2}.   
The average stress tensor for a microcanonical ensemble of strings with total energy $E$ is related to the on-shell thermalon value by \cite{Mertens et al 2}
\begin{equation}
\langle T^{\mu \nu}_{e}\rangle_{E} \propto T^{\mu \nu}_{\mathrm{th}}\mid _{\mathrm{on-shell}}
\end{equation}
where

\begin{equation}
\langle T^{\mu \nu}_{e}\rangle_{E} \approx \frac{2}{\sqrt{G}}e^{-{\beta}_H E } \frac{\delta}{\delta G_{\mu \nu}} (\frac{e^{{\beta}_H E}}{{\beta}_H})
\end{equation}

The thermalon description gives a good approximation where the energy of the condensate is large, which will be the case if the kinetic energy of collapse is converted into string, as proposed in \cite{Hewitt2014} . The thermalon also gives an effective description of the average string charge density, with the $U(1)$ thermalon charge used in the superconductor model of \cite{Hewitt2014} related to the string charge current by \cite{Mertens et al 2}
 \begin{equation}
J^{\mu \nu}_e = -\frac{2}{\sqrt{G}}\frac{\delta S_e}{\delta B_{\mu \nu}} 
\end{equation}
and
\begin{equation}
J^{\tau k}_e \propto J^k _{\mathrm{on-shell}}
\end{equation}

\section{Thermalon weak field solutions}
These are relevant to the beginning of the proposed conversion process, and show that there is no energy barrier to initiating conversion.  For small values of the thermal string density we can neglect interactions \cite{Hewitt2014} \cite{Mertens et al 1}.

The heterotic thermalon field equation in Rindler coordinates is \cite{Hewitt2014}:
\begin{displaymath}
\frac{\partial^2 \phi}{\partial r^2} + \frac{1}{r}\frac{\partial \phi}{\partial r} = m^2(\beta (r)) \phi = (\frac{1}{4r^2}+ 4r^2 -6)\phi \label{DE}
\end{displaymath}
where the thermalon mass $m$ is effectively $r$ dependent:
\begin{displaymath}
m^2(r) = (\frac{1}{4r^2} +4r^2 -6)
\end{displaymath}
This admits the following finite accelerating wall solution
\begin{displaymath}
\phi =\epsilon \sqrt{r} \exp (-r^2) \label{sol}
\end{displaymath}
where $\epsilon$ is a small parameter.\\

The area difference across the solution is given by
\begin{equation}
2\pi \delta A \sim 4\pi GA \int_{0}^{\infty}\mu \beta dr = 2\pi GA(\frac{\pi}{2})^{3/2}\epsilon^2
\end{equation}

This can be described by the warp factor $w$ where

\begin{equation}
w-1 = G (\frac{\pi}{2})^{3/2}\epsilon^2
\end{equation} 

Type II strings show similar behaviour \cite{Mertens et al 1}.
and in particular the energy density is given by

\begin{equation}
\langle T^{\tau \tau}_{e} (\textbf{x}) \rangle _E = -\frac{2E}{\sqrt{\alpha '}}(2-\frac{\alpha '}{\rho ^2})e^{-\rho^2 / \alpha '}
\end{equation}
which is negative for

\begin{displaymath}
\rho < \sqrt{\frac{\alpha '}{2}}
\end{displaymath}
with positive radial pressure
\begin{equation}
\langle T^{\rho \rho}_e (\textbf{x}) \rangle _E = \frac{2E}{\sqrt{\alpha '}}e^{-\rho^2 / \alpha '}
\end{equation}
for the type II thermalon solution
\begin{equation}
\psi \sim \epsilon e^{-\rho^2 / \alpha '}
\end{equation}
thus having the qualitative behaviour found for heterotic strings \cite{Hewitt2014}, \cite{Mertens et al 1} that leads to the possibility of a self-supporting condensate around a collapsed object \cite{Hewitt2015}, \cite{Mertens et al 3}.
This shows that contrary to the expectation expressed in \cite{Hewitt2014}, thermal duality not is not directly required for a self-supporting condensate to form. For type II strings, the shell thickness is $O(1)$ rather than $ \log{M}$, and no interaction is required for a self-supporting condensate. Further investigation may show that this is the same in the heterotic case, rather than the superconducting ball picture of \cite{Hewitt2014} . In this case, the string condensate shell has a conical rather than hyperbolic geometry. The volume of the interior is still proportional to area for the final state, with entropy proportional to area. Our conjecture also implies that (some generalisation of) string or brane condensation also provides a gravitational regulator for 11 dimensional supergravity.\\

 \section{Gravito-thermal interaction}
For heterotic strings the left moving currents and their interaction vertex, at the dual point, the pair  (thermalon, graviton) is equivalent to $(W, \gamma )$ The thermalon can be seen as part of the gravity sector in string theory generally, and gravity becomes part of a  combined `gravito-thermal' interaction. The condensate that forms at the Hagedorn transition behaves as part of the gravity sector as it can be described by the thermalon. A string condensate may form near black hole \cite{Hewitt2014} \cite{Mertens et al 3} or possibly replace it \cite {Hewitt2015}  This condensate is self-supporting due to negative energy on the inside \cite{Hewitt2015} \cite{Mertens et al 2}, allowing the possibility of a conversion process as the interior shrinks \cite{Hewitt2014}. The existence of such states augmented by a coating of string condensate is a key stepping stone to the proposed conversion process. For such a condensate to form, we propose that it is not necessary to gently introduce static energy at the Hagedorn point near a black hole as might be expected, but rather that there is an automatic conversion of matter in free fall to produce such a string condensate, and further that the final state will be essentially composed of this condensate without a closed trapping horizon. A demonstration of the conversion effect through concrete calculation is the main missing link in the programme proposed here. \\

This effect would  limit the strength of gravity, preventing the formation of black holes, which would be a novel macroscopic effect. String theory in 4 dimensions is problematic without a gravity regulator \cite{Hewitt2015}. With a regulator, it may become like QCD, with gravitationally free strings confined to a bag of a different phase \cite{Hewitt2015}. We will show that the proposed gravity regulator is generic in spacetime dimension, and potentially avoids black hole formation and related information paradoxes in types of string theory.\\

\section{Gravity Regulator proposal}
 The basic proposal is that the acceleration of convex surfaces with locally constant area is limited by the Hagedorn Rindler acceleration.  Note that this condition is not related to the magnitude of the tidal Riemannian curvature. \\

This adiabatic (constant area) condition may be expressed as follows: Let $E_{0}$ the timelike unit vector field be a Killing vector and be tangent to the boundary hypersurface $B$ then the area 2 form $\sigma$ within $B$ is, with $\phi$ the embedding of $B$ into spacetime $M$
\begin{equation}
\sigma = *_{B} \phi ^{*}( e^{0})
\end{equation}
and the condition at the initial boundary $I$ of $B$ for $B$ to have constant area is, using an overdot to denote the Lie derivative $\mathcal{L}_{E_0}$
\begin{equation}
\dot \sigma = \ddot \sigma= 0
\end{equation}
so that on $B$ we will have $\dot \sigma = 0$, or
\begin{equation}
i_{E_0}d*_{B} e^0 = i_{E_0}d\sigma = 0
\end{equation}
by Cartan's magic formula.\\

The critical acceleration condition is now
\begin{equation}
i_{E_0} De^0 = D_{E_0} e^0 = 2\pi T * \Sigma
\end{equation}
where $\Sigma$ is the volume 3-form of $B$, so that $*\Sigma$ is normal to $B$. \\

For nucleation to begin at $I$ it is also necessary that $I$ be convex so that the warp factor can increase as nucleation progresses. In terms of the extrinsic curvature of $I$,
\begin{equation}
\chi_{a b} =g^{\mu \nu}*\Sigma^{\mu} D_{(a} D_{b)} *\Sigma_{\nu} 
\end{equation}
this requires
\begin{equation}
\mathrm{Tr}( \chi) > 0 , \: \: \mathrm{det}(\chi) >0
\end{equation}
and $I$ has positive curvature by the $\textit{therorema egregium}$. In particular, these conditions prevent the nucleation process from occurring in free space, and nucleation is restricted to the environment of collapsed or collapsing objects.\\

 Note the generic nature of the conditions, which  may apply across M space as the differential geometry applies in all uncompactified dimensions and a self-supporting condensate is known for different string types, although the mechanism for heterotic strings was conjectured on the basis of the self-dual nature of the thermalon \cite{Hewitt2014}. 
The critical surface conditions are a string regularised version of a Penrose trapped surface (which is the infinite string tension limit for these surfaces). Horizons are defined globally but trapped surfaces (and thermalon traps) are defined locally by these geometric conditions.\\

 Heuristically the critical gravity condition for conversion is equivalent to the condition that the Newtonian gravity (in local centre of mass frame) between any two objects is limited by the string tension \cite{Hewitt2015}. This equivalence is universal in that it applies for any uncompactified spacetime dimension. The mutual nature of the proposed novel gravito-thermal effect between two interacting bodies is shown by the symmetry of the Newtonian effective criterion for the gravitational regulator, which may hold throughout M space. \\

Similarly to the Newtonian force limit, the conversion power, the rate at which the kinetic energy of an infalling body is converted to string condensate, would take a universal value related to the string tension in any spacetime dimension. This is consistent with the conjecture \cite{Hewitt2015} that the decelerating body generates thermalons in a way analogous to the Larmor formula for the power emitted by an accelerated charge during the conversion process .

\section{Transition process}
A thermalon shell region is stable on the outside, unstable/dynamic on the inside. Assume that thermalon polarization maintains gravitational regulation throughout gravitational collapse. Transition would begin at a separation $\sim \sqrt{Mm}$ between objects of mass  $M$ and $m$. Deceleration may couple matter to thermalon production to give a stringy bremsstrahlung process, converting kinetic energy to thermalons, in a dynamic aspect of the gravito-thermal interaction. The conversion power is constant $\sim T$ (the string tension) throughout relative to local static frames. This is consistent with conversion being a gravity sector effect. Note that conversion will very quickly become thermodynamically favoured over continued free fall once the geometric criteria are met. The conversion process is dissipated and can be tracked through time. The string condensate acts as a recording medium which can capture the information carried by the initial state. The entropy in this model arises from entanglement between short and long string sectors \cite{Hewitt2014} and increases during conversion from the point of view of an observer with access only to the short string sector. In this way, the conversion process would be on the same footing as any other dissipative physical process.\\

Relativistic factors for infalling matter can be estimated as $\gamma \sim M $ for initial formation and  $\gamma \sim d$ for particle absorption, where $d$ is the distance from the existing boundary. This gives estimates for transition times (as seen from infinity):  $\sim M^{d-3} \log{M}$  for the initial formation, and  $\sim M^{d-3} \log{\sqrt{Mm}}$ for particle absorption.\\

Is conversion to string condensate from Minkowski space possible? Suppose a planar thermalon trap crosses a wall of matter in Minkowski space. The wall of matter cannot be converted to a thermalon excitation because of the warp factor that would be produced. This agrees with the expectation that this process should be forbidden by the conservation of energy. \\

Consider next a thermalon front geometry with a salient. Could this be produced by a thermalon trap absorbing a particle in Minkowski space? An expanding wave-front bounds a convex region in a Minkowski background, which is inconsistent with it having a locally constant area. The area of the trailing surface cannot decrease as this is flat and already has minimum area. Thus, although traps associated with Rindler frames are to be found everywhere in Minkowski space, they cannot gain energy by recruiting particles that they encounter.\\

\section{Hot holograms}
The final state in this scenario is in thermal equilibrium with the exterior normal vacuum (up to a slow escape of radiation equivalent to the Hawking process). A hot object with matching temperature and surface gravity can be in thermal equilibrium with normal vacuum, with  $T_{\alpha \beta} =0$ in the exterior space. In general, the surface temperature and gravity would only match (i.e. $2 \pi T = g$)  by coincidence and the object would be unstable.
A thermalon deformation however naturally and stably meets this matching condition for temperature and surface gravity as it is tied to the location of high static gravitational acceleration. The collapsed object is stable against collapse because the interior region has undergone a radial collapse and there is nowhere left for the hot string to collapse into, because of their essentially 2 dimensional nature, these final states can be thought of as hot holograms formed from the collapsing matter content \cite{Hewitt2015}, and would constitute  'deep firewalls' in the terminology of \cite{Braunstein Pirandola}. \\

\section{Resolution of paradoxes?}
This model may resolve the `firewall paradoxes'  by removing the horizon, and providing a physically motivated firewall to store the information content of a collapsing object. The `firewall' here would be formed by string bremsstrahlung during gravitational collapse. The firewall would be stable  - there is no black hole, and the interior space has been crushed so there is nowhere else for the energy and information to go except outward by Hawking evaporation. Because of condensation, string models may have a vacuum polarization mechanism which limits the strength of gravity. The criterion for this to be effective is based on the area of accelerating surfaces, not on the magnitude of the tidal Riemannian curvature. According to our conjecture, this effect would occur at, and only at, sites of extreme gravitational collapse. Collapsing objects would form hot holograms, rather than black holes. This would resolve the firewall paradox in the following way: firewalls would be real, but black holes would not. 

\section{Acknowledgements}
M.H. wishes to thank George Leontaris for his kind invitation to present this work in Ioannina, and Nick Mavromatos for his encouragement.

\end{document}